\documentclass[twocolumn,showpacs,preprintnumbers,amsmath,amssymb,prb,aps]{revtex4}

\usepackage{epsfig}% Include figure files
\usepackage{graphicx}% Include figure files
\usepackage{dcolumn}% Align table columns on decimal point
\usepackage{bm}% bold math

\begin{document}

\title{Correlation induced half-metallicity in a ferromagnetic single-layered compound: Sr$_2$CoO$_4$}
\author{Sudhir K. Pandey}

\altaffiliation{Electronic mail: sk$_{_{-}}$iuc@rediffmail.com}

\affiliation{UGC-DAE Consortium for Scientific Research, University
Campus, Khandwa Road, Indore - 452001, India}

\date{\today}

\begin{abstract}
The electronic and magnetic properties of Sr$_2$CoO$_4$ compound
have been studied using $\emph{ab initio}$ electronic structure
calculations. As opposed to GGA calculation, which gives
ferromagnetic metallic solution, GGA+$U$ calculations provide two
kind of ferromagnetic solutions: (i) half-metallic and (ii)
metallic. The half-metallic solution is a ground state of the system
and the metallic one is a metastable state. The strong hybridization
between Co 3$d$ and O 2$p$ orbitals decides the electronic and
magnetic properties of the compound. The total magnetic moment per
formula unit is found to be $\sim$ 3 $\mu_B$ ($S$ = 3/2). Our
calculations give the magnetocrystalline anisotropy energy of $\sim$
2.7 meV, which provides a good description of experimentally
observed large magnetocrystalline anisotropy. The Heisenberg
exchange parameters up to fourth nearest neighbours are also
calculated. The mean-field theory gives the $T_C$ = 887 K. The
possible physical implications of the ferromagnetic half-metallic
ground state are also discussed.

\end{abstract}


\maketitle

\section{Introduction}
The single-layered compounds with a general formula of A$_2$BO$_4$
(A and B stand for rare-earth/alkali metals and transition metals,
respectively) have attracted a great deal of attention after the
discovery of high temperature superconductor in
La$_{2-x}$Sr$_x$CuO$_4$.\cite{imada} These compounds show many
exotic physical phenomena like spin/charge stripes formation in
nickelates and manganites\cite{imada}; and spin-triplet
superconductivity in ruthenates.\cite{maeno}

Among the single-layered compounds recently synthesized
Sr$_2$CoO$_4$ has been reported to show ferromagnetic (FM) and
metallic behaviours, which have never been found in any other such
materials.\cite{matsuno,wang} The magnetization data show $T_C$
around 250 K. The large magnetic anisotropy where magnetic easy axis
is the $c$-axis is also observed. Moreover, the resistivity data
reveal the quasi-two-dimensional electronic nature for the system.
Interestingly, there are contradicting reports on the magnetic
moments. The Matsuno $\emph{et al}$.\cite{matsuno} and Wang $
\emph{et al}$.\cite{wang} have reported the saturation magnetization
of about 1.8 and 1.0 $\mu_B$/Co, respectively. The effective
magnetic moment of 3.72 $\mu_B$ has also been found by fitting
Curie-Wiess behaviour in the paramagnetic phase, which suggests the
magnetic moment corresponds to $S$ $\approx$ 3/2 spin
configuration.\cite{wang}

Lee $\emph{et al}$.\cite{lee} tried to understand the above
contradictory reports on the saturation magnetization by using
LDA+$U$ calculations. Their studies suggest that such values of
magnetization can be intrinsic to the system as they got similar two
values for the total magnetic moment for different ranges of $U$.
However, if we read the works of Matsuno $\emph{et al}$. and Wang
$\emph{et al}$. carefully we can easily make out that the different
values of saturation magnetization they got may not be intrinsic to
the system. The M(H) data of both the groups look similar. For
example, the values of total magnetization at 5 T found by Wang
$\emph{et al}$. and Matsuno $\emph{et al}$. are about 1.4 and 1.6
$\mu_B$/Co, respectively, which are not very different. The
different values of saturation magnetization quoted in the last
paragraph appear to arise due to two different approaches the
authors used in extracting them.

The partial replacement of Sr by rare-earth elements leads to the
formation of interesting magnetic phases keeping the crystal
structure intact.\cite{shimada,ang,huang} For example,
Sr$_{1.5}$La$_{0.5}$CoO$_4$ compound retains the FM ground state of
Sr$_2$CoO$_4$, whereas SrLaCoO$_4$ manifests spin glass
state.\cite{shimada} Contrary to FM state for
Sr$_{1.5}$La$_{0.5}$CoO$_4$, the ground state of
Sr$_{1.5}$Pr$_{0.5}$CoO$_4$ compound is found to be spin
glass.\cite{ang} Similarly, the formation of spin glass state is
also reported in Sr$_{1.25}$Nd$_{0.75}$CoO$_4$ compound.\cite{huang}
The existence of spin glass state in the doped compounds appears to
be a generic phenomenon. It is well known that the spin glass state
arises due to the presence of competing ferromagnetic and
antiferromagnetic interactions. Therefore, it would be interesting
to calculate the nature of magnetic interactions between the
neighbouring Co atoms in the Sr$_2$CoO$_4$ compound.

Here, we report the detailed electronic and magnetic states of
Sr$_2$CoO$_4$ using $\emph{ab initio}$ electronic structure
calculations. The ground state solution is found to be a
ferromagnetic half-metallic state having $S$ = 3/2 spin
configuration, which is different from that reported by Lee
$\emph{et al}$. The strong hybridization between Co 3$d$ and O 2$p$
orbitals decides the electronic and magnetic properties of the
compound. The calculation gives magnetocrystalline anisotropy energy
of $\sim$ 2.7 meV. The nature of magnetic interactions between the
Co atoms is calculated up to fourth nearest neighbours, which shows
mixed ferromagnetic and antiferromagnetic coupling. The estimated
ferromagnetic transition temperature within mean-field theory is
$\sim$ 887 K.

\section{Computational details}
The nonmagnetic and ferromagnetic electronic structure calculations
of Sr$_2$CoO$_4$ compound were carried out using LmtArt
6.61.\cite{savrasov} For calculating charge density, full-potential
linearized Muffin-Tin orbital (FP-LMTO) method working in plane wave
representation was employed. In the calculations, we have used the
Muffin-Tin radii of 2.915, 1.965, 1.608, and 1.778 a.u. for Sr, Co,
O1, and O2, respectively. The charge density and effective potential
were expanded in spherical harmonics up to $l$ = 6 inside the sphere
and in a Fourier series in the interstitial region. The initial
basis set included 5$s$, 4$p$, and 4$d$ valence, and 4$s$ semicore
orbitals of Sr; 4$s$, 4$p$, and 3$d$ valence, and 3$p$ semicore
orbitals of Co, and 2$s$ and 2$p$ orbitals of O. The exchange
correlation functional of the density functional theory was taken
after Vosko {\em et al}.\cite{vosko} and GGA calculations were
performed following Perdew {\em et al}.\cite{perdew}

The effect of on-site Coulomb interaction ($U$) under GGA+$U$
formulation of the density functional theory is also considered in
the calculations. The detailed description of the GGA+$U$ method
implemented in the code can be found in Ref. [12]. The double
counting scheme used in the code is normally called as fully
localized limit in the literature. To study the role of orbital
degrees of freedom, we have also included spin-orbit coupling (SOC)
in the calculations. The Self-consistency was achieved by demanding
the convergence of the total energy to be smaller than 10$^{-5}$
Ry/cell. (10, 10, 10) divisions of the Brillouin zone along three
directions for the tetrahedron integration were used to calculate
the density of states (DOS).

\begin{figure}
  \includegraphics[width=8.5cm]{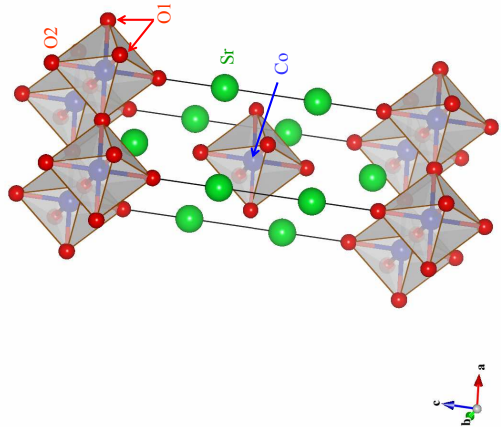}\\
  \caption{(Color online) Atomic arrangements in the unit cell. Sr, Co
and O atoms are represented by spheres with decreasing
radii.}\label{Fig1}
\end{figure}

\section{Results and discussions}
The atomic arrangement in the unit cell is shown in Fig. 1, which
displays bodycenter tetragonal lattice. The Co atoms occupy the
corners and bodycenter positions and each Co atom is surrounded by
six O atoms forming a distorted octahedron. In this structure there
are two kinds of O represented by O1 and O2. O1 lies in the $ab$
plane and O2 along the $c$-axis. The bond distance between the
corner and bodycenter Co atoms is almost 1.8 times larger than that
between nearest Co atoms sitting at the corners. This suggests the
quasi-two-dimensional nature of the system where its electronic
properties are expected to be decided by transport of electrons in
the $ab$ plane.

\begin{figure}
  \includegraphics[width=8.5cm]{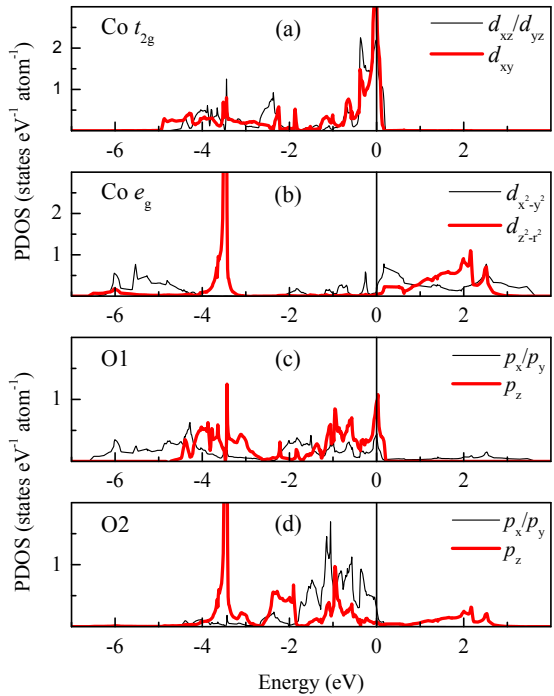}\\
  \caption{(Color online) The partial density of different symmetric
states obtained from nonmagnetic solution. (a) $d_{xz}/d_{yz}$ (thin
lines) and $d_{xy}$ (thick lines) states, (b) $d_{x^2-y^2}$ (thin
lines) and $d_{z^2-r^2}$ (thick lines) states. The $p_x/p_y$ (thin
lines) and $p_z$ (thick lines) symmetric partial density of states
of planar oxygen (O1) and apical oxygen (O2) are shown in (c) and
(d), respectively.}\label{Fig2}
\end{figure}

In order to know the crystal-field effect and nature of Co-O bonding
in Sr$_2$CoO$_4$ we have plotted the partial density states (PDOS)
of Co 3$d$ and O 2$p$ obtained from nonmagnetic solution in Fig. 2.
In the octahedral symmetry Co 3$d$ states split into $t_{2g}$ and
$e_g$ states and the separation between these states is found to be
$\sim$ 1.5 eV. The degeneracies of these states are further lifted
in tetragonal symmetry as evident from Figs. 2(a) and (b). The
triply degenerate $t_{2g}$ states split into doubly degenerate
($d_{xz}$, $d_{yz})$ states and nondegenerate $d_{xy}$ state.
Similarly, doubly degenerate $e_g$ states split into nondegenerate
$d_{x^{2}-y^{2}}$ and $d_{z^{2}-r^{2}}$ states. The O 2$p$ orbitals
also split in doubly degenerate ($p_x$, $p_y$) orbitals and
nondegenerate $p_z$ orbital as evident from Figs. 2(c) and (d).

The $t_{2g}$ and $e_g$ sectors are spread over the energy range of
about 5 and 10 eV, respectively. Roughly two times larger extent of
the $e_g$ states is due to larger overlap between $e_g$ and $p$
orbitals in comparison to that between $t_{2g}$ and $p$ orbitals. In
the present situation one can get maximum overlaps of
$d_{x^{2}-y^{2}}$ and O1 ($p_x$, $p_y$) orbitals; and
$d_{z^{2}-r^{2}}$ and O2 $p_z$ orbitals which can lead to larger
bandwidth of $e_g$ symmetric states. The occupied $d_{z^{2}-r^{2}}$
states are mostly found in the narrow region of 1 eV whereas
occupied $d_{x^{2}-y^{2}}$ states are spread over an energy window
of about 4 eV. Total number of $d$ electrons is found to be $\sim$
6.5, which is about 1.5 more than the expected nominal value for
Co$^{4+}$ ion. The above observations can be considered as a
signature of the covalent nature of the Co-O bonds. There are large
Co 3$d$ PDOS at the $E_F$ ($\sim$ 8.4 states/eV/atom) in the
nonmagnetic solution which may be considered as a signature of
ferromagnetic ground state based on the Stoner theory.

\begin{figure}
  \includegraphics[width=8.5cm]{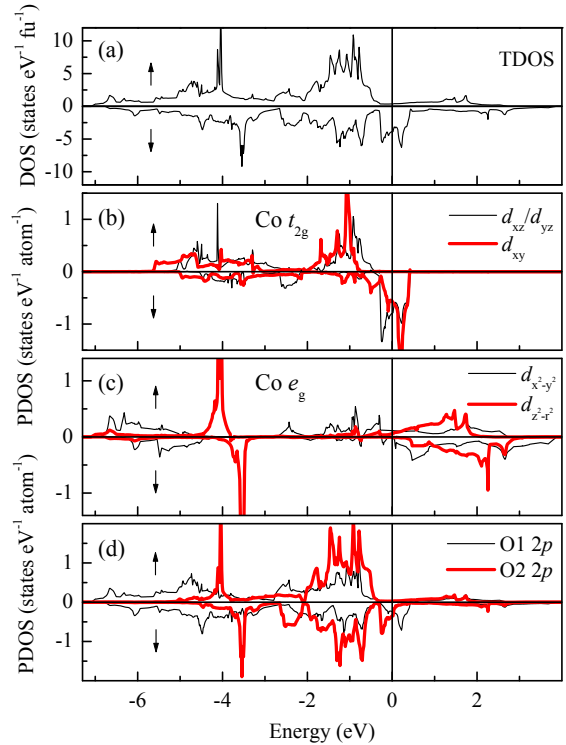}\\
  \caption{(Color online) The density of states obtained from
ferromagnetic GGA calculation. (a) total density of states (TDOS),
(b) $d_{xz}/d_{yz}$ (thin lines) and $d_{xy}$ (thick lines) states,
and (c) $d_{x^2-y^2}$ (thin lines) and $d_{z^2-r^2}$ (thick lines)
states. In (d) partial density of $p$ symmetric states for  planar
oxygen (O1) and apical oxygen (O2) are represented by thin and thick
lines, respectively.}\label{Fig3}
\end{figure}

In order to know the electronic and magnetic structures of the
compound we have carried out FM calculation. The energy of the FM
solution is found to be $\sim$ 336 meV less than that of nonmagnetic
solution indicating the ferromagnetic ground state for the system.
This result is in accordance with the experimental observation and
earlier theoretical calculations.\cite{matsuno,lee} The total and
partial density of states obtained for the FM configuration are
plotted in Fig. 3. There is an asymmetric distribution of states in
both the spin channels. The difference between number of electrons
in up and down spin channels is a measure of net magnetization and
found to be $\sim$ 2. Under the rigid band picture one can estimate
the effective value of the exchange parameter ($J$) by considering
the energy difference between the up and down band edges, which is
estimated to be $\sim$ 0.4 eV. The $t_{2g}$ up-spin bands are fully
occupied whereas $t_{2g}$ down-spin and $e_g$ bands are partially
occupied.

Similar to the nonmagnetic case, $e_g$ bands are highly extended in
comparison to $t_{2g}$ bands. The strength of exchange parameters
for $t_{2g}$ and $e_g$ electrons are estimated to be $\sim$ 0.6 and
0.4 eV, respectively. The larger value of $J$ for $t_{2g}$ electrons
can be attributed to more localized $t_{2g}$ states leading to
higher Coulomb interaction. Interestingly, O 2$p$ PDOS also show
spin polarization and the estimated value of $J$ for $p$ electrons
is also found to be $\sim$ 0.4 eV. The values of magnetic moment for
Co, O1 and O2 atoms are found to be about 1.52, 0.17 and 0.07
$\mu_B$, respectively. The total magnetization per formula unit is
$\sim$ 2 $\mu_B$, which is almost same to the earlier
reports.\cite{matsuno,lee} The appearance of magnetic moments at the
O sites may be attributed to the strong hybridization between
$d_{x^{2}-y^{2}}$ and O1 ($p_x$, $p_y$) orbitals; and
$d_{z^{2}-r^{2}}$ and O2 $p_z$ orbitals. The values of magnetic
moment for O1 and O2 atoms also appear to support this conjecture as
the magnetic moment of O1 is roughly 2 times larger than that of O2.

In order to know the role of orbital degrees of freedom on the
magnetic properties of the system we have included SOC in the
calculations by considering magnetization directions along (001) and
(100) axes. Our calculations indicate that the $c$-axis is an easy
axis as the total energy of the system for (001) direction is found
to be $\sim$ 2.7 meV/Co less than that for (100) direction. This
result is in line with the experimentally observed magnetic easy
axis.\cite{matsuno} This value of magnetocrystalline anisotropy
energy (MAE) is $\sim$ 2.1 meV/Co more than that obtained by Lee
$\emph{et al}$.\cite{lee} Based on the magnetization data of Ref.
[3], Lee \emph{et al}. have given a rough estimate of $\sim$ 1.5
meV/Co for the MAE. This value is about three times larger than that
of the Lee \emph{et al}. and about half of our value. The inclusion
of SOC keeps the spin magnetic moments almost intact and induces 0.1
$\mu_B$ orbital magnetic moment at the Co site.

The above GGA calculations provide the good representation of
experimentally observed electronic and magnetic properties of the
system including magnetocrystalline anisotropy. However, the on-site
Coulomb interaction parameter $U$ has been found to be important in
understanding the detailed electronic and magnetic properties of the
3$d$ transition metals oxies.\cite{anisimov} Therefore, it would be
interesting to see the effect of $U$ on electronic and magnetic
states of the compound under GGA+$U$. In this approximation $U$ and
$J$ are used as parameters.

The values of $U$ and $J$ required to reproduce the electronic
structure of a compound are sensitive to the approximations used in
the calculations.\cite{sudhirLCO,ong} Moreover, it is well known
that the GGA+$U$ calculations often converge to the local minima
depending on the starting electronic configurations
used.\cite{korotin,knizek,dorado} In the light of these facts we
have varied $U$ from 3 to 5 eV and fixed the value of $J$ to 0.4 eV
as estimated for the $e_g$ electrons. To know the exact ground state
we have also used different starting configurations for the Co 3$d$
electrons. This range of $U$ is found to provide the good
representation of the experimentally observed electronic and
magnetic properties of different Co based
systems.\cite{sudhirLCO,sudhirPCO,sudhirIPES,wu1,wu2}

\begin{figure}
  \includegraphics[width=8.5cm]{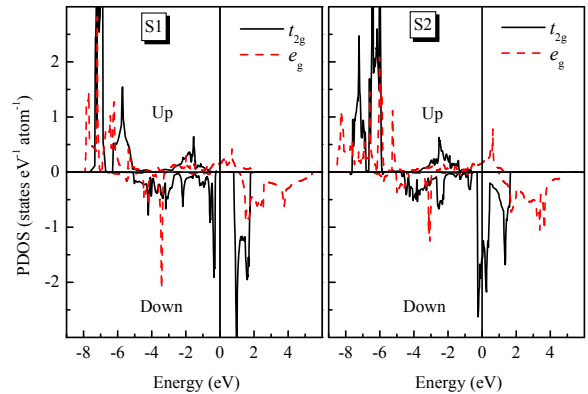}\\
  \caption{(Color online) The partial density of Co 3$d$ symmetric
states obtained from ferromagnetic GGA+$U$ ($U$ = 4 eV and $J$ = 0.4
eV) calculations corresponding to two conversed solutions
represented by S1 and S2. $t_{2g}$ and $e_g$ orbitals are denoted by
solid and dashed lines, respectively.}\label{Fig4}
\end{figure}

Based on symmetry consideration we used 10 different starting
electronic configurations for the Co atom in the calculations. For
these configurations only two kind of FM solutions S1 and S2 are
found to exist. The S1 and S2 give half-metallic and metallic
states, respectively. The difference in the electronic structure of
Co in both the solutions can be found from Fig. 4, where we have
plotted the Co 3$d$ PDOS calculated for $U$ = 4 eV. The
half-metallicity with a gap of about 1 eV in the down-spin channel
is evident for the S1. The S2 shows metallic state with a large
$t_{2g}$ symmetric states around the $E_F$ in the down-spin channel.
In the up-spin channel both the solutions provide similar PDOS of
$e_g$ character in the vicinity of $E_F$. The comparison of Co 3$d$
PDOS obtained from the S2 and FM GGA solutions indicates that S2 is
originating from the FM GGA solution. The S1 solution is quite
different and it has origin to the orbital polarization.

The PDOS closer to $E_F$ along with occupancies of the Co 3$d$
orbitals listed in Table 1 can be used to characterize both the
solutions. Numbers written in normal and italics are corresponding
to the S1 and S2, respectively. The $t_{2g}$$^{\uparrow}$ states are
fully occupied in both the solutions. The main difference between
the two solutions arises from the $t_{2g}$$^{\downarrow}$ sector. In
the S1 and S2, $d_{xy}$$^{\downarrow}$ state can be considered as
fully occupied and unoccupied, respectively. The contribution of
$d_{xz}$$^{\downarrow}$/$d_{yz}$$^{\downarrow}$ state below the
$E_F$ is small ($\sim$ 0.17 electron) and can be considered as fully
unoccupied in the S1, which gives rise to the half-metallic state.
However, $d_{xz}$$^{\downarrow}$/$d_{yz}$$^{\downarrow}$ state is
roughly half-filled ($\sim$ 0.6) in the S2 and mainly responsible
for the metallic state. Thus, we can represent the S1 and S2 by
$d_{xy}$$^{\downarrow1}$ and
$d_{xz}$$^{\downarrow1/2}$/$d_{yz}$$^{\downarrow1/2}$ electronic
configurations, respectively.

Table 1: Electronic occupancies of different Co 3$d$ orbitals along
with the total number of $d$ electrons for both the spin channels
obtained from ferromagnetic GGA+$U$ ($U$ = 4 eV and $J$ = 0.4 eV)
calculations corresponding to two self consistent solutions
represented by S1 (normal) and S2 (italics).

\vspace{2ex}
\begin{ruledtabular}
\begin{tabular}{|c|c|c|c|c|c|}
    & $d_{xz}$/$d_{yz}$ & $d_{xy}$ & $d_{x^2-y^2}$ & $d_{z^2-r^2}$ & Total $d$\\
  \hline
  Up & 0.95,$\emph{0.94}$ & 0.94,$\emph{0.95}$ & 0.84,$\emph{0.90}$ & 0.88,$\emph{0.82}$ & 4.56,$\emph{4.55}$ \\
  Down & 0.17,$\emph{0.61}$ & 0.92,$\emph{0.14}$ & 0.22,$\emph{0.25}$ & 0.36,$\emph{0.25}$ & 1.84,$\emph{1.86}$ \\
  \hline
  Up - Down & 0.78,$\emph{0.33}$ & 0.02,$\emph{0.81}$ & 0.62,$\emph{0.65}$ & 0.52,$\emph{0.57}$ & 2.72,$\emph{2.69}$ \\
\end{tabular}
\end{ruledtabular}

Both the solutions give roughly same magnetic moment ($\sim$ 2.7
$\mu_B$) for the Co. The S1 induces magnetic moment of $\sim$ 0.08
$\mu_B$ at the O2 site whereas the S2 provides $\sim$ 0.18 $\mu_B$
at the O1 site. The total magnetic moment per formula unit for both
the solutions is found to be $\sim$ 3 $\mu_B$/fu representative of
$S$ = 3/2 spin state. Interestingly, the energy of the S1 is $\sim$
234 meV less than that of the S2 indicating that the S1 corresponds
to the true ground state of the system. Present work clearly shows
the importance of $U$ in stabilizing a half-metallic state which
turns out to be the lowest energy state of the system. Thus, the
ground state of Sr$_2$CoO$_4$ is found to be half-metallic with $S$
= 3/2 spin configuration.

Our half-metallic ground state is quite different from that obtained
in earlier work.\cite{lee} The half-metallic solution of Lee
$\emph{et al}$. gives $S$ = 1/2 spin configuration. This solution
corresponds to almost occupied
$d_{xz}$$^{\downarrow}$/$d_{yz}$$^{\downarrow}$ states and
half-filled $d_{xy}$$^{\downarrow}$ state whereas our S1 corresponds
to the opposite situation where
$d_{xz}$$^{\downarrow}$/$d_{yz}$$^{\downarrow}$ states are almost
vacant and $d_{xy}$$^{\downarrow}$ state is filled. Here it is worth
mentioning that we have also included the generalized gradient
approximation in the calculations whereas Lee \emph{et al}. have
only considered the local density approximation (LDA). This may give
an impression that the different formulations of the exchange
correlation functional can be the reason for the difference between
our results and that of the Lee \emph{et al}. To address this issue
we have also carried out detailed calculations using spin polarized
LDA+$U$ method. The results obtained from these calculations are
found to be almost the same as those obtained by GGA+$U$. This
indicates that the existence of two solutions has nothing to do with
the approximations involved in treating different exchange
correlation functional and solely related to the orbital
polarization.

\begin{figure}
  \includegraphics[width=8.5cm]{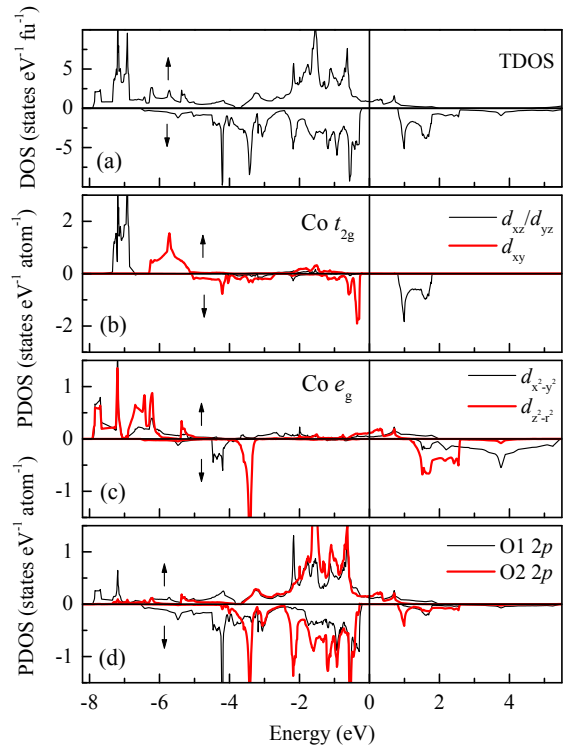}\\
  \caption{(Color online) The density of states corresponds to the
ferromagnetic half-metallic solution obtained from GGA+$U$ ($U$ = 4
eV and $J$ = 0.4 eV) calculation. (a) total density of states
(TDOS), (b) $d_{xz}/d_{yz}$ (thin lines) and $d_{xy}$ (thick lines)
states, and (c) $d_{x^2-y^2}$ (thin lines) and $d_{z^2-r^2}$ (thick
lines) states. In (d) partial density of $p$ symmetric states for
planar oxygen (O1) and apical oxygen (O2) are represented by thin
and thick lines, respectively.}\label{Fig5}
\end{figure}

The total and partial densities of states correspond to the
half-metallic solution are shown in Fig. 5. The large spin
polarization is evident from the total density of states plotted in
Fig. 5(a). The lowest occupied band shows the signature of van Hove
singularity arising from the $e_g$ symmetric states as evident from
Fig. 5 (c). The $d_{xy}$ orbital is fully occupied and a gap of
$\sim$ 1 eV is created between it and the unoccupied
$d_{xz}$/$d_{yz}$ orbitals in the down-spin channel. In the up-spin
channel only $e_g$ sector provides the conducting electrons with a
dominating contribution from the $d_{z^2-r^2}$ electrons. The
$d_{x^2-y^2}$ symmetric PDOS in the vicinity of the $E_F$ is almost
flat suggesting that the temperature dependent behaviour of the
compound should be decided by $d_{z^2-r^2}$ electrons. Fig. 5(b)
shows a huge exchange splitting ($\sim$ 7.5 eV) for the degenerate
$d_{xz}$/$d_{yz}$ orbitals. This result also suggests that the
Coulomb correlation in the presence of strong hybridization between
Co 3$d$ and O 2$p$ orbitals along with a tetragonal distortion makes
full occupation of nondegenerate $d_{xy}$ orbital energetically more
favorable. This may be considered as a signature of a
ferrodistortive Jahn-Teller ordering.

In order to study the detailed nature of magnetic interactions we
have calculated the Heisenberg exchange parameters $J_{0i}$ between
the Co atoms using FLEUR code,\cite{fleur} where 0 and $i$ stand for
the central atom and coordination shell index, respectively. The Co
atom sitting at the origin is considered as the central atom, Fig.
1. The values of $J_{0i}$ up to 4$^{th}$ nearest neighbours
(coordination shells) are shown in Table 2. It is evident from the
table that the first and third nearest neighbour interactions are
ferromagnetic whereas second and fourth nearest neighbour
interactions are antiferromagnetic in nature. The value of $J_{01}$
is found to be 19.06 meV which is about 15 times larger than the
values of neighbouring interactions. This suggests that the
effective magnetic interaction in Sr$_2$CoO$_4$ is ferromagnetic in
nature, which is in accordance with the experimentally observed
ferromagnetic ground state.\cite{matsuno} Under the mean-field
approximation we calculated the Curie temperature ($T_C$) of $\sim$
887 K, which is roughly 3.5 times larger than the experimentally
measured value. Such discrepancy between the experimental and
calculated $T_C$ is not surprising as the overestimation of $T_C$
under mean-field approximation is well known in the
literature.\cite{liechtenSSC,liechtenJMMM,shi}

Table 2: The Heisenberg exchange parameters ($J_{0i}$) up to fourth
nearest neighbours (coordination shells) obtained by ferromagnetic
GGA calculations using FLEUR code. 0 and $i$ stand for central atom
and coordination shell index, respectively.

\vspace{2ex}
\begin{ruledtabular}
\begin{tabular}{|c|c|c|}
  Nearest neighbour & Coordination number & $J_{0i}$ (meV) \\
  \hline
  1$^{st}$ & 4 & 19.06 \\
  2$^{nd}$ & 4 & -1.11 \\
  3$^{rd}$ & 8 & 1.25 \\
  4$^{th}$ & 4 & -0.33 \\

\end{tabular}
\end{ruledtabular}

Finally, we discuss the physical implications of our results. The
half-metallic ferromagnet has potential application in the
sprintonics industry.\cite{wolf} This compound can be considered as
a good candidate as it shows $T_C$ closer to the room temperature.
The $d$ PDOS at the $E_F$ are quite low (0.19 states/eV/atom) as
evident from the S1 of Fig. 4. This suggests that even a moderate
hole doping at the Co sites can lead to insulating ground state. The
present work predicts the saturation magnetization of $\sim$ 3
$\mu_B$/fu. The 5K M(H) data of Wang $\emph{et al}$. do not show any
sign of saturation at maximum measured field of 5 T where the value
of magnetization is found to be about 1.4 $\mu_B$/Co.\cite{wang}
However, their M(T) data suggest to support our calculated $S$ = 3/2
spin state as they have reported the effective magnetic moment of
$\sim$ 3.72 $\mu_B$. Here it is important to note that M(H) data of
Matsuno $\emph{et al}$. show almost saturation of magnetization at 7
T where the value of magnetization is found to be about 1.8
$\mu_B$/Co.\cite{matsuno} However, this work has been carried out on
the thin films, which may suffer from the effects of substrate
induced strain and/or defects present in the films. Moreover,
substrate contribution to the magnetization and the weight of the
material in the thin film samples are difficult to measure
accurately. All these things may lead to different magnetization
value than bulk sample. Interestingly, Matsuno $\emph{et al}$. have
explained their saturation magnetization value by assuming Co$^{4+}$
in intermediate spin state and that would lead to the same magnetic
moment we obtained.

\section{Conclusions}
The detailed electronic and magnetic properties of a single-layered
compound - Sr$_2$CoO$_4$ - have been studied using $\emph{ab
initio}$ calculations with different formulations of the density
functional theory. The GGA calculation gives a ferromagnetic
metallic state with total magnetic moment of $\sim$ 2 $\mu_B$/fu.
The GGA+$U$ calculations provide two ferromagnetic solutions with
$S$ = 3/2 spin configuration: (i) half-metallic and (ii) metallic.
The ferromagnetic half-metallic state is the lowest energy state and
representative of the ground state. Our results also suggest that
the compound may show metal to insulator transition when a small
amount hole is doped at the Co sites. In the present work we have
also estimated the crystal-field energy, magnetic moments,
Heisenberg exchange parameters, transition temperature,
$\emph{etc}$., which provide good description of experimentally
observed electronic and magnetic properties of the system.

%\section{Acknowledgements}

\end{document}